\documentclass[aps,pra,twocolumn,superscriptaddress,showpacs]{revtex4-1}
\usepackage{graphicx}
\begin{document}
[Phys. Rev. A {\bf 82}, 063637 (2010)]
\title{Polarized entangled Bose-Einstein condensation}
\author{Jinlong Wang}
\affiliation{State Key Laboratory of Surface Physics and  Department of Physics, Fudan University, Shanghai,
200433, China}
\author{Yu Shi}
\email{ yushi@fudan.edu.cn}
\affiliation{State Key Laboratory of Surface Physics and  Department of Physics, Fudan University, Shanghai,
200433, China}
\affiliation{Department of Physics, The University of Texas, Austin, TX 78712}

\begin{abstract}
We consider a mixture of two distinct species of atoms of  pseudospin-$\frac{1}{2}$ with both intraspecies and Interspecies spin-exchange interactions, and find all the ground stats in a general case of the parameters in the effective Hamiltonian. In general, corresponding to the two species and two pseudo-spin states, there are four orbital wave functions into which the atoms condense. We find that in certain parameter regimes, the ground state is the so-called  polarized entangled  Bose-Einstein condensation, i.e. in addition to condensation of interspecies singlet pairs, there are unpaired atoms with spins polarized in the same direction. The interspecies entanglement and polarization significantly affect the generalized Gross-Pitaevskii equations governing the four orbital wave functions into which the atoms condense, as an interesting interplay between spin and orbital degrees of freedom.
\end{abstract}
\pacs{03.75.Mn, 03.75.Gg}
\maketitle
\section{introduction}

Since a decade ago, there have been a lot of activities on
Bose-Einstein condensation (BEC) of atoms with  spin degree of freedom~\cite{BECD}, for example, in spin-1~\cite{ho1,CKL,ueda,TLH4031,spinor} and  pseudospin-$\frac{1}{2}$ gases~\cite{al,ABK170403,SA063612,li}, as well as BEC  of a mixture of two kinds of spinless atoms~\cite{BECD,homix,pu,ao,esry,timmermans,trippenbach,myatt}. As a step beyond these research lines, a novel class of BEC, the so-called entangled BEC (EBEC), was proposed as the ground state of a mixture of two distinct species of pseudospin-$\frac{1}{2}$ atoms with interspecies spin exchange in some parameter regime~\cite{shi0,shi1,shi2}. Spin-exchange scattering between atoms of different species entangles these two species, and EBEC may lead to a novel kind of superfluidity~\cite{shi3}. More recently, EBEC has also been found in a mixture of spin-$1$ atoms~\cite{shi4}. It has been noted that  EBEC may be experimentally realized using two different species of alkali atoms. In some aspects, EBEC bears some analogies with a single species of pseudospin-$\frac{1}{2}$ atoms in a double well~\cite{al} or occupying two orbital modes~\cite{ABK170403,SA063612}, but there are also important differences due to the fact that two atoms of distinct species are distinguishable.   In general,  there are four orbital wave functions into which the atoms of  the  two species and two pseudospin states condense into, respectively.

Under single-mode approximation for orbital degree of freedom, the many-body Hamiltonian can be written as~\cite{shi1,shi2,shi3}
\begin{widetext}
\begin{equation}
\mathcal{H}=\sum_{\alpha,\sigma}
f_{\alpha\sigma}N_{i\sigma}+\frac{1}{2}\sum_{\alpha,\sigma\sigma'}
K^{(\alpha\alpha)}_{\sigma\sigma'}N_{\alpha\sigma}N_{\alpha\sigma'}
+\sum_{\sigma\sigma'}
K^{(ab)}_{\sigma\sigma'}N_{a\sigma}N_{b\sigma'}
+K_e (a^\dagger_\uparrow
a_\downarrow b^\dagger_\downarrow b_\uparrow+a^\dagger_\downarrow
a_\uparrow b^\dagger_\uparrow b_\downarrow), \label{h}
\end{equation}
\end{widetext}
where $\alpha_{\sigma}$ ($\alpha = a, b$, $ \sigma =\uparrow, \downarrow$) is the annihilation operator of the species $\alpha$,
\begin{equation}
K^{(\alpha\beta)}_{\sigma\sigma'}\equiv
g^{(\alpha\beta)}_{\sigma\sigma'}\int
\phi_{\alpha\sigma}^*(\mathbf{r})\phi_{\beta\sigma'}^*(\mathbf{r})
\phi_{\beta\sigma'}(\mathbf{r})\phi_{\alpha\sigma}(\mathbf{r})
d^3r
\end{equation}
is related to the scattering lengths as the following. In general, $\xi^{(\alpha\beta)}_{\sigma_1\sigma_2\sigma_3\sigma_4}$ is
the scattering length for the scattering in which an $\alpha$-atom
flips from $\sigma_4$ to $\sigma_1$ while a $\beta$-atom flips from
$\sigma_3$ to $\sigma_2$,
$\mu_{\alpha\beta}=m_{\alpha}m_{\beta}/(m_{\alpha}+m_{\beta})$ is
the reduced mass.
Now, for intraspecies scattering,  $g^{(\alpha\alpha)}_{\sigma\sigma }\equiv
2\pi\hbar^2\xi^{(\alpha\alpha)}_{\sigma\sigma\sigma\sigma}
/\mu_{\alpha\alpha}$ for same species and same spin, $g^{(\alpha\alpha)}_{\sigma\bar{\sigma} }\equiv
4\pi\hbar^2\xi^{(\alpha\alpha)}_{\sigma\bar{\sigma}\bar{\sigma}\sigma}
/\mu_{\alpha\alpha}= 4\pi\hbar^2\xi^{(\alpha\alpha)}_{\sigma\bar{\sigma}\sigma\bar{\sigma}}
/\mu_{\alpha\alpha}$ for $\sigma \neq \bar{\sigma}$.  For interspecies scattering without spin exchange, $g^{(ab)}_{\sigma\sigma'}\equiv
2\pi\hbar^2\xi^{(ab)}_{\sigma\sigma'\sigma'\sigma}/\mu_{\alpha\beta}$, where $\sigma$ and $\sigma'$ may or may not be equal. For interspecies scattering with spin exchange,
\begin{equation} K_e \equiv g_e \int
\phi_{a\sigma}^*(\mathbf{r})\phi_{b\bar{\sigma}}^*(\mathbf{r})
\phi_{b\sigma}(\mathbf{r})\phi_{a\bar{\sigma}}(\mathbf{r})
d^3r,
\end{equation}
where $\sigma \neq \bar{\sigma}$, $g_e \equiv g^{(ab)}_{\sigma\bar{\sigma}\sigma\bar{\sigma}} \equiv
2\pi\hbar^2\xi^{(ab)}_{\sigma\bar{\sigma}\sigma\bar{\sigma}}
/\mu_{\alpha\beta}$.
$f_{\alpha\sigma} \equiv \epsilon_{\alpha\sigma}-
K^{(\alpha\alpha)}_{\sigma\sigma}/2,$ where $\epsilon_{\alpha\sigma}
= \int \phi_{\alpha\sigma}^*h_{\alpha\sigma} \phi_{\alpha\sigma}
d^3r $ is the single-particle energy, $h_{\alpha\sigma}$ being the single-particle Hamiltonian.

The total spin operator of species $\alpha$ ($\alpha=a,b$)
is $\mathbf{S}_{\alpha}=
\alpha^\dagger_\sigma\mathbf{s}_{\sigma\sigma'}\alpha_\sigma'$, where $\mathbf{s}_{\sigma\sigma'}$ is the single spin operator.
Hence the Hamiltonian can be transformed into that of two coupled giant spins,
\begin{widetext}
\begin{equation}
\mathcal{H}=2K_e(S_{ax}S_{bx}+S_{ay}S_{by})+J_zS_{az}S_{bz}+B_a S_{az}+B_b S_{bz}+C_aS_{az}^2+C_bS_{bz}^2+E_0,
\end{equation}
\end{widetext}
where
$J_z=K_{\uparrow\uparrow}^{(ab)}+K_{\downarrow\downarrow}^{(ab)}
-K_{\uparrow\downarrow}^{(ab)}-K_{\downarrow\uparrow}^{(ab)},$
$B_a=f_{a\uparrow}-f_{a\downarrow}+\frac{N_a}{2}(K_{\uparrow\uparrow}^{(aa)}
-K_{\downarrow\downarrow}^{(aa)})+\frac{N_b}{2}(K_{\uparrow\uparrow}^{(ab)}
+K_{\uparrow\downarrow}^{(ab)} -K_{\downarrow\uparrow}^{(ab)}
-K_{\downarrow\downarrow}^{(ab)}),$
$B_b=f_{b\uparrow}-f_{b\downarrow}+\frac{N_b}{2}(K_{\uparrow\uparrow}^{(bb)}
-K_{\downarrow\downarrow}^{(bb)})+
\frac{N_a}{2}(K_{\uparrow\uparrow}^{(ab)}
+K_{\downarrow\uparrow}^{(ab)}
-K_{\uparrow\downarrow}^{(ab)}-K_{\downarrow\downarrow}^{(ab)}),$
$C_a = \frac{1}{2}(K_{\uparrow\uparrow}^{(aa)}
+K_{\downarrow\downarrow}^{(aa)}-K_{\uparrow\downarrow}^{(aa)}
-K_{\downarrow\uparrow}^{(aa)})$,
$C_b =\frac{1}{2}(K_{\uparrow\uparrow}^{(bb)}
+K_{\downarrow\downarrow}^{(bb)}-K_{\uparrow\downarrow}^{(bb)}
-K_{\downarrow\uparrow}^{(bb)}).$ $E_0=(1/2)\sum_{\alpha}(\sum_{\sigma}
f_{\alpha\sigma})N_{\alpha}+(1/8)\sum_{\alpha}(\sum_{\sigma\sigma'}
K_{\sigma\sigma'}^{(\alpha\alpha)})
N_{\alpha}^2+(1/4)(\sum_{\sigma\sigma'}K_{\sigma\sigma'}^{(ab)})N_aN_b$ does not depend on spins. We label the two species in such a way that $N_a \geq N_b$.

As discussed previously, at the antiferromagnetic isotropic point of the effective parameters, namely $2K_e=J_z>0$ while $B_a=B_b=C_a=C_b=0$,
$\mathcal{H}= 2 K_e \mathbf{S}_a\cdot\mathbf{S}_b$. Then the ground states are $
|\frac{N_a-N_b}{2}, S_z\rangle = Z(a_{\uparrow}^{\dagger})^{N_a/2-N_b/2+S_z}
(a_{\downarrow}^{\dagger})^{N_a/2-N_b/2-S_z}
(a_{\uparrow}^{\dagger}b_{\downarrow}^{\dagger}-
a_{\downarrow}^{\dagger}b_{\uparrow}^{\dagger})^{N_b}|0\rangle,$
where $Z$ is the normalization constant. This ground state leads to interesting physical consequences. It has been shown numerically and analytically that EBEC also persists in a considerable regime away from the isotropic point.

In this paper, we extend the consideration of this model to a more general case of the parameters, namely,  $B_a=B_b$,  denoted as $B$, $C_a = C_b$, denoted as $C$, and $J_z-2K_e-2C = 0$, and thus  the Hamiltonian becomes
\begin{equation}
\mathcal{H}= 2 K_e \mathbf{S_a}\cdot\mathbf{S_b}+BS_z+C S_z^2+E_0. \label{hamiltonianall}
\end{equation}
which reduces to the isotropic antiferromagnetic coupling when $B=C=0$ while $K_e >0$.  In other words, we assume that  there is such a rotational symmetry  that the Hamiltonian only depends on the total $S$ and $S_z$, which are thus good quantum numbers. We believe that under a perturbation due to deviation of the parameters from this condition, the ground state is still close to the present one in each parameter regime discussed below, based on an extension of the previous analysis for the isotropic parameter point~\cite{shi2}, the details of which will be reported elsewhere.

In Sec.~\ref{gss}, we find all the ground states of the Hamiltonian (\ref{hamiltonianall}) in various parameter regimes, based on the calculation accounted in the Appendices. In Sec.~\ref{boseexp}, these  ground states are expressed in terms of Bosonic degrees of freedom. Most of the ground states are of the form of $|S,\pm S\rangle$, which are referred to as the polarized BEC if $S\neq 0$, as then  $S_z=\pm S \neq 0$. Their properties are discussed in Sec~\ref{prop}.  A summary is given in Sec.~\ref{summary}.

\section{Ground States in Various parameter regimes \label{gss}}

For spin-$\frac{1}{2}$ atoms,
$S_a=N_a/2$ and $S_b=N_b/2$ are fixed. It is clear that the total spin $S$ and its $z$-component $S_z$ are good quantum numbers, hence the eigenstates are of the form of $|S,S_z\rangle$. The ground state is thus
\begin{equation}
|G\rangle = |S^m,S_z^m\rangle, \label{gs} \end{equation} where  ${\cal S}^m$ and  ${\cal S}_z^m$ are, respectively, the values of $S$ and $S_z$ that minimize
\begin{equation}
E  =  K_e S(S+1) + B S_z + C S_z^2 \label{e1}
\end{equation}
In case $C\neq 0$, it can be rewritten as
\begin{equation}
E  =  K_e\left(S+\frac{1}{2}\right)^2+
C\left(S_z+\frac{B}{2C}\right)^2+E_0',
\label{e2}
\end{equation}
where $E_0'$ is independent of $S$ and  $S_z$. The minimization should be under the constraints
\begin{eqnarray}
S_{min}  \leq & S & \leq S_{max}, \label{range1}  \\
-S     \leq & S_z & \leq S, \label{range2}
\end{eqnarray}
where
\begin{equation}
S_{min} \equiv S_{a}-S_{b} = \frac{N_a-N_b}{2}, \label{smin}
\end{equation}
\begin{equation}
S_{max} \equiv S_{a}+S_b = \frac{N_a+N_b}{2}.
\end{equation}

We have obtained all the ground states in the whole three-dimensional $C-K_e-B$ parameter space.

\subsection{$B\neq 0$}

\begin{figure*}
\scalebox{1.0}[1.0]{\includegraphics{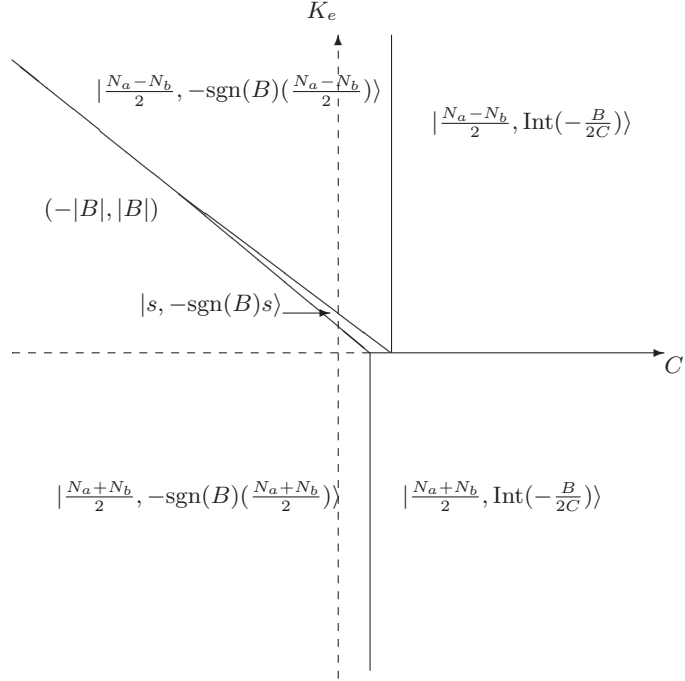}}
\caption{\label{pd1} Ground states in $C-K_e$ parameter plane for a given $B\neq 0$, which can be positive or negative. The solid lines are boundaries between different regimes. The dashed lines are only coordinate axes. The ground states in five regimes are indicated.  $|s,-{\rm sgn}(B)s\rangle$, with $s \equiv {\rm Int}(\frac{|B|-K_e}{2(K_e+C)})$, is the ground state in the regime surrounded by  $K_e=\frac{|B|-(N_a+N_b)C}{N_a+N_b+1}$, $K_e = \frac{|B|-(N_a-N_b)C}{N_a-N_b+1}$ and $K_e=0$ , which is continuously connected with $|\frac{N_a+N_b}{2},-{\rm sgn}(B)(\frac{N_a+N_b}{2})\rangle$  and $|\frac{N_a-N_b}{2},-{\rm sgn}(B)(\frac{N_a-N_b}{2})\rangle$ on the two boundaries, respectively.  The boundary for $C < -|B|$ is  $K_e =  \frac{|B|-N_aC}{N_a+1}$, on which both
$|\frac{N_a+N_b}{2},-{\rm sgn}(B)(\frac{N_a+N_b}{2})\rangle$  and $|\frac{N_a-N_b}{2},-{\rm sgn}(B)(\frac{N_a-N_b}{2})\rangle$ are degenerate ground states. Three boundary lines converge at $K_e=-C=|B|$, on which any state is the ground state. The ground state $|\frac{N_a+N_b}{2},-{\rm Int}(\frac{B}{2C})\rangle$ is continuously connected with $|\frac{N_a+N_b}{2},-{\rm sgn}(B)(\frac{N_a+N_b}{2})\rangle$ on the boundary   $C = \frac{|B|}{N_a+N_b}$ while $K_e < 0$.
$|\frac{N_a-N_b}{2}, {\rm Int}(-\frac{B}{2C})\rangle$ is continuously connected with $|\frac{N_a-N_b}{2}, -{\rm sgn}(B)(\frac{N_a-N_b}{2})\rangle$ on the boundary  $C  = \frac{|B|}{N_a-N_b}$ while $K_e>0$.  On the boundary $K_e=0$ while $\frac{|B|}{N_a+N_b} \leq C \leq \frac{|B|}{N_a-N_b}$, the degenerate ground states are  of the form of $|S^m, {\rm Int}(-\frac{B}{2C})\rangle$ with  $|{\rm Int}(-\frac{B}{2C})| \leq S^m \leq \frac{N_a+N_b}{2}$. On the boundary $K_e=0$ while $C \geq \frac{|B|}{N_a-N_b}$, the degenerate ground states are  of the form of $|S^m, {\rm Int}(-\frac{B}{2C})\rangle$ with  $\frac{N_a-N_b}{2} \leq S^m \leq \frac{N_a+N_b}{2}$. This figure reduces to FIG.~\ref{pdn} in the special case of $N_a=N_b$.  }
\end{figure*}

\begin{figure*}
\scalebox{1.2}[1.2]{\includegraphics{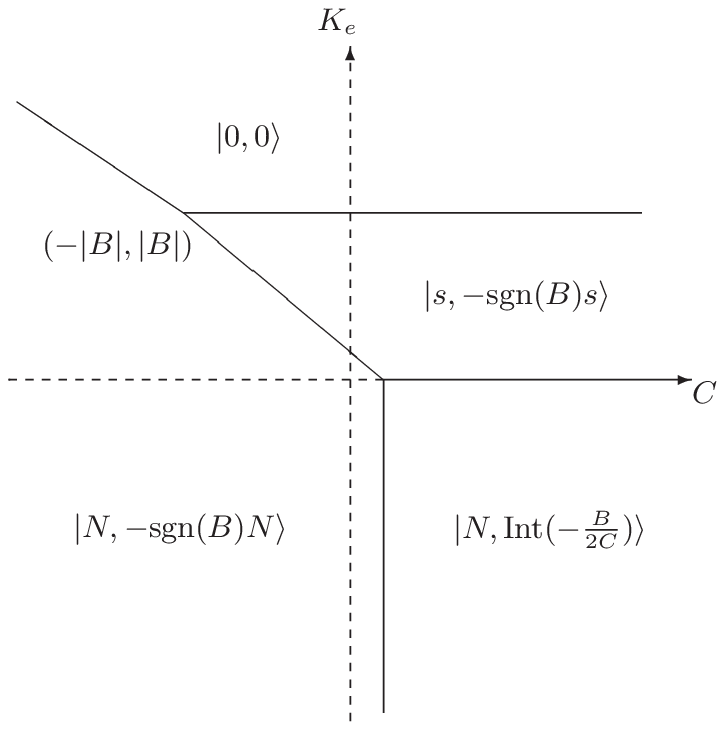}}
\caption{\label{pdn} Ground states in $C-K_e$ parameter plane for a given $B\neq 0$ in the special case of $N_a=N_b=N$. This diagram is a limit  of FIG.~\ref{pd1}, with the boundary $C=\frac{|B|}{N_a-N_b}$ disappearing  while the boundary $K_e=\frac{|B|-(N_a-N_b)C}{N_a-N_b+1}$ becoming a horizontal line $K_e=|B|$. }
\end{figure*}

Based on the calculations in Appendix~\ref{bneq0}, the ground states for  $B\neq 0$ are summarized in Table~\ref{table1}, and  in the  two-dimensional $C-K_e$ phase diagrams for a given  $B\neq 0$, as shown in FIG.~\ref{pd1}, as well as FIG.~\ref{pdn} for the special case of $N_a=N_b=N$.

\begin{table*}
\begin{tabular}{|l|l|l|l|}
\hline \multicolumn{3}{|c|}{Parameter regimes} & {Ground states $|S^m,S_z^m\rangle$} \\
\hline
 &   &$ K_e < \frac{|B|-N_a C}{N_a+1}$  &$|\frac{N_a+N_b}{2},-{\rm sgn}(B)(\frac{N_a+N_b}{2})\rangle$ \\ \cline{3-4}
 &$C < -|B|$ & $K_e=\frac{|B|-N_a C}{N_a+1}$ &$|\frac{N_a-N_b}{2},-{\rm sgn}(B)(\frac{N_a-N_b}{2})\rangle$ and \\
 & & & $|\frac{N_a+N_b}{2},-{\rm sgn}(B)(\frac{N_a+N_b}{2})\rangle$ \\ \cline{3-4}
 & & $K_e>\frac{|B|-N_a C}{N_a+1}$ &$|\frac{N_a-N_b}{2},-{\rm sgn}(B)(\frac{N_a-N_b}{2})\rangle$ \\ \cline{2-4}
 & & $K_e < -C$ & $|\frac{N_a+N_b}{2},-{\rm sgn}(B)(\frac{N_a+N_b}{2})\rangle$ \\ \cline{3-4}
 & $C=-|B|$ & $K_e=-C$ & $|S^m, -{\rm sgn}(B)S^m\rangle$, $\frac{N_a-N_b}{2} \leq S^m \leq \frac{N_a+N_b}{2}$ \\ \cline{3-4}
 & & $K_e > -C$ & $|\frac{N_a-N_b}{2},-{\rm sgn}(B)(\frac{N_a-N_b}{2})\rangle$ \\ \cline{2-4}
 &  &$K_e \leq \frac{|B|-(N_a+N_b)C}{N_a+N_b+1}$ &$|\frac{N_a+N_b}{2},-{\rm sgn}(B)(\frac{N_a+N_b}{2})\rangle$ \\ \cline{3-4}
 $B\neq 0$&$-|B| < C\leq \frac{|B|}{N_a+N_b}$ &$\frac{|B|-(N_a+N_b)C}{N_a+N_b+1}\leq K_e\leq\frac{|B|-(N_a-N_b)C}{N_a-N_b+1}$ &$|s,-{\rm sgn}(B)s\rangle$, $s={\rm Int}(\frac{|B|-K_e}{2(K_e+C)})$  \\ \cline{3-4}
 & &$K_e \geq \frac{|B|-(N_a+N_b)C}{N_a+N_b+1}$  &$|\frac{N_a-N_b}{2},-{\rm sgn}(B)(\frac{N_a-N_b}{2})\rangle$ \\ \cline{2-4}
 & & $K_e < 0$ & $|\frac{N_a+N_b}{2}, {\rm Int}(-\frac{B}{2C})\rangle$ \\ \cline{3-4}
 & & $K_e=0$ & $|S^m, {\rm Int}(-\frac{B}{2C})\rangle$, $|{\rm Int}(-\frac{B}{2C})| \leq S^m \leq \frac{N_a+N_b}{2}$ \\ \cline{3-4}
 & $\frac{|B|}{N_a+N_b} \leq C \leq \frac{|B|}{N_a-N_b}$ & $0 < K_e \leq \frac{|B|-(N_a-N_b)C}{N_a-N_b+1}$ & $|s, -{\rm sgn}(B)s\rangle$, $s={\rm Int}(\frac{|B|-K_e}{2(K_e+C)})$  \\ \cline{3-4}
 & & $K_e \geq \frac{|B|-(N_a-N_b)C}{N_a-N_b+1}$ & $|\frac{N_a-N_b}{2}, -{\rm sgn}(B)\frac{N_a-N_b}{2}\rangle$ \\ \cline{2-4}
 & & $K_e < 0$ & $|\frac{N_a+N_b}{2}, {\rm Int}(-\frac{B}{2C})\rangle$ \\ \cline{3-4}
 &$C\geq\frac{|B|}{N_a-N_b}$  &$K_e=0$ &$|S^m, {\rm Int}(-\frac{B}{2C})\rangle$, $|{\rm Int}(-\frac{B}{2C})| \leq S^m \leq \frac{N_a+N_b}{2}$ \\ \cline{3-4}
 & & $K_e > 0$ & $|\frac{N_a-N_b}{2}, {\rm Int}(-\frac{B}{2C})\rangle$ \\ \hline
\end{tabular}
\caption{\label{table1} Ground states in various parameter regimes with $B\neq 0$.}
\end{table*}

In the regime $K_e <  \frac{|B|-N_aC}{N_a+1}$ for $C < -|B|$ plus  $K_e \leq \frac{|B|-(N_a+N_b)C}{N_a+N_b+1}$ for $-|B|< C \leq \frac{|B|}{N_a+N_b}$, the ground state is $|\frac{N_a+N_b}{2},-{\rm sgn}(B)(\frac{N_a+N_b}{2})\rangle$, where ${\rm sgn}(B)$ denotes the sign of $B$.

In the regime $ K_e > \frac{|B|-N_aC}{N_a+1}$ for $C < -|B|$ plus $K_e \geq \frac{|B|-(N_a-N_b)C}{N_a-N_b+1}$ for $-|B|<C \leq \frac{|B|}{N_a-N_b} $, the ground state is $|\frac{N_a-N_b}{2},-{\rm sgn}(B)(\frac{N_a-N_b}{2})\rangle$. Note that if $N_a=N_b$, then this statement is still valid by regarding  $\frac{|B|}{N_a-N_b}$ as infinity. In other words,
the boundary $C = \frac{|B|}{N_a-N_b}$  disappears, and the ground state is $|0,0\rangle$ in the whole regime of $K_e > \frac{|B|-N_aC}{N_a+1}$ for $C < -|B|$ plus $K_e \geq |B|$ for $C > -|B|$.

On the boundary $ K_e = \frac{|B|-N_aC}{N_a+1}$ for $C < -|B|$, $|\frac{N_a+N_b}{2},-{\rm sgn}(B)(\frac{N_a+N_b}{2})\rangle$ and $|\frac{N_a-N_b}{2},-{\rm sgn}(B)(\frac{N_a-N_b}{2})\rangle$ are the two degenerate ground states.

In the regime  $\frac{|B|-(N_a+N_b)C}{N_a+N_b+1} \leq K_e  \leq \frac{|B|-(N_a-N_b)C}{N_a-N_b+1}$ and $K_e>0$ for  $-|B|< C \leq \frac{|B|}{N_a-N_b}$,  the ground state is $|s,-{\rm sgn}(B)s\rangle$, where $s \equiv {\rm Int}(S_0)$ is the integer closest to $S_0 \equiv \frac{|B|-K_e}{2(K_e+C)}$ and  in the legitimate range $S_{min} \leq s \leq S_{max}$.   When $N_a=N_b$, the boundary $K_e= \frac{|B|-(N_a-N_b)C}{N_a-N_b+1}$ becomes $K_e=|B|$ for $C> -|B|$ up to infinity.
This is a crossover regime, as $s$ depends on the parameters, and is continuously connected with the two neighboring regimes, that is, $s=\frac{N_a+N_b}{2}$ at the boundary $K_e =\frac{|B|-(N_a+N_b)C}{N_a+N_b+1}$ while $s=\frac{N_a-N_b}{2}$ at the boundary $K_e =\frac{|B|-(N_a-N_b)C}{N_a-N_b+1}$.
But there are discontinuities on the boundary $K_e=0$ for  $\frac{|B|}{N_a+N_b}\leq  C\leq \frac{|B|}{N_a-N_b}$.

In the regime $C \geq \frac{|B|}{N_a+N_b}$ while $K_e < 0$,  the ground state is $|\frac{N_a+N_b}{2},-{\rm Int}(\frac{B}{2C})\rangle$.
It is continuously connected with
$|\frac{N_a+N_b}{2},-{\rm sgn}(B)(\frac{N_a+N_b}{2})\rangle$  on the boundary $C=\frac{|B|}{N_a+N_b}$ while $K_e\leq 0$.

Finally, for $N_a > N_b$, in the regime $C \geq \frac{|B|}{N_a-N_b}$ while $K_e>0$, the ground state is $|\frac{N_a-N_b}{2}, {\rm Int}(-\frac{B}{2C})\rangle$. It is continuously connected with $|\frac{N_a-N_b}{2},-{\rm sgn}(B)(\frac{N_a-N_b}{2})\rangle$ on the boundary $C=\frac{|B|}{N_a-N_b}$ while $K_e >0$.

On the boundary $K_e=0$ while  $\frac{|B|}{N_a+N_b} \leq C \leq \frac{|B|}{N_a-N_b}$,  the degenerate ground states are $|S^m, {\rm Int}(-\frac{B}{2C})\rangle$, with $|{\rm Int}(-\frac{B}{2C})| \leq S^m \leq \frac{N_a+N_b}{2}$. On the boundary $K_e=0$ while  $C \geq \frac{|B|}{N_a-N_b}$, the degenerate ground states are $|S^m, {\rm Int}(-\frac{B}{2C})\rangle$, with  $\frac{N_a-N_b}{2} \leq S^m \leq \frac{N_a+N_b}{2}$.  The ground state in each regime bordering each of these two boundaries is one of the ground states on it. Hence discontinuities or quantum phase transitions occur in crossing this boundary from one regime to another.

The only other  boundary where discontinuity or quantum phase transitions occur is $ K_e = \frac{|B|-N_aC}{N_a+1}$ for $C < -|B|$. In entering it from downside or upside, the ground state remains as the original one, but when leaving it and entering the other side, the ground state discontinuously changes to another state. Nevertheless, if entering the boundary from the crossover regime $\frac{|B|-(N_a+N_b)C}{N_a+N_b+1} \leq K_e  \leq \frac{|B|-(N_a-N_b)C}{N_a-N_b+1}$ and $K_e >0$ for  $-|B|< C \leq \frac{|B|}{N_a-N_b}$, there is no discontinuity, as the ground state $|s,-{\rm sgn}(B)s\rangle$ in this regime is continuously connected with both $|\frac{N_a+N_b}{2},-{\rm sgn}(B)(\frac{N_a+N_b}{2})\rangle$ and $|\frac{N_a-N_b}{2},-{\rm sgn}(B)(\frac{N_a-N_b}{2})\rangle$.

Three  boundaries $ K_e = \frac{|B|-N_aC}{N_a+1}$ for $C < -|B|$, $K_e= \frac{|B|-(N_a+N_b)C}{N_a+N_b+1}$ and $K_e = \frac{|B|-(N_a-N_b)C}{N_a-N_b+1}$ for  $C > -|B|$
converge at $K_e=-C=|B|$, which does not belong to any of these three regimes and  where the ground state can be any state $|S^m,S_z^m\rangle$,  with $S^m$ and $S_z^m$ in the legitimate ranges, or any superposition of these states.
In the three dimensional $C-K_e-B$ parameter space, these three boundaries that are lines on the $C-K_e$ plane with a given $B$ become planes. $K_e=-C=|B|$ are two lines converging at the origin for $B \geq 0$ and $B \leq 0$ respectively.

The structure of the three-dimensional phase diagram possesses a reflection symmetry with respect to the plane $B=0$.  In each pair of  symmetric regimes, the ground states have the same $S^m$ but opposite $S_z^m$.

As $B \rightarrow 0$, the boundary $C = \frac{|B|}{N_a+N_b}$  for $K_e \leq 0$ approaches the negative half of $K_e$-axis including the origin, while the boundary $C = \frac{|B|}{N_a-N_b}$  for $K_e > 0$, existing when $N_a > N_b$, approaches the positive half of $K_e$-axis excluding the origin. The boundary  $ K_e = \frac{|B|-N_aC}{N_a+1}$  for $C < -|B|$ approaches $ K_e = \frac{-N_aC}{N_a+1}$  for $C < 0$ , while the crossover regime of  $|s,-{\rm sgn}(B)s\rangle$ tends to vanish, and $K_e=-C=|B|$ approaches the origin $K_e=-C=|B|=0$.

\subsection{$B=0$}

\begin{figure*}
\scalebox{1.2}[1.2]{\includegraphics{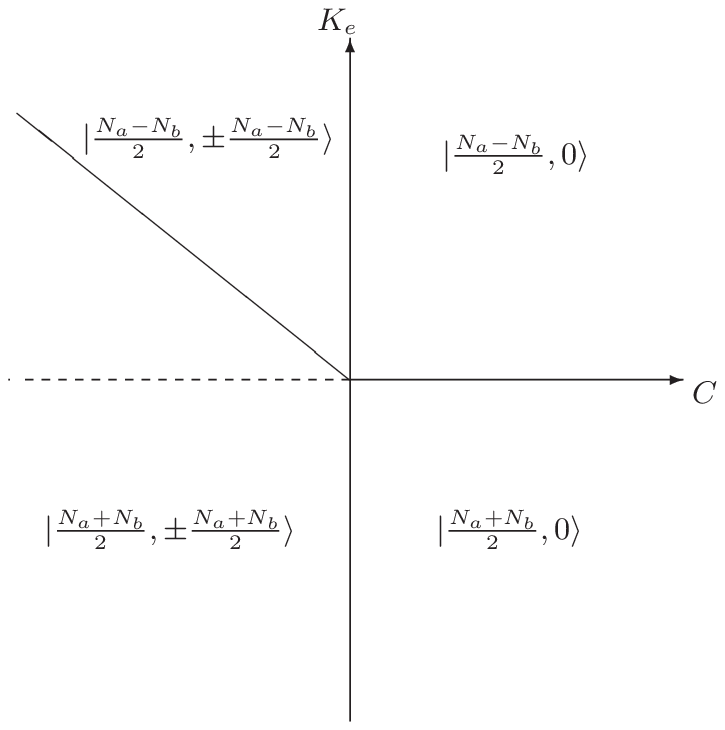}}
\caption{\label{pd2} Ground states in $C-K_e$ parameter plane for  $B=0$. The solid lines are boundaries between different regimes. The dashed line is only the negative $C$ axis.   The ground states in four regimes are indicated. The boundary for $C < 0$ is  $K_e =  -\frac{N_aC}{N_a+1}$, on which both
$|\frac{N_a+N_b}{2},\pm \frac{N_a+N_b}{2}\rangle$  and $|\frac{N_a-N_b}{2},\pm \frac{N_a-N_b}{2}\rangle$ are degenerate ground states. On the positive $C$ axis, the ground states are $|S^m,0\rangle$, with $\frac{N_a-N_b}{2} \leq S^m \leq  \frac{N_a+N_b}{2}$.   On the positive $K_e$ axis, the ground states are $|\frac{N_a-N_b}{2},S^m_z\rangle$, with $-\frac{N_a-N_b}{2} \leq S^m_z \leq \frac{N_a-N_b}{2}$. On the negative $K_e$ axis, the ground states are $|\frac{N_a+N_b}{2},S^m_z\rangle$, with  $-\frac{N_a+N_b}{2} \leq S^m_z \leq \frac{N_a+N_b}{2}$. On the origin
$B=C =K_e= 0$, the ground state can be any superposition of $|S^m,S_z^m\rangle$, with  $\frac{N_a-N_b}{2} \leq S^m \leq  \frac{N_a+N_b}{2}$ and $-S^m \leq S^m_z \leq  S^m $.  }
\end{figure*}

Based on the calculations in Appendix~\ref{beq0}, the ground states  for $B=0$ are summarized in Table~\ref{tableb0}, and depicted in the  two-dimensional $C-K_e$ phase diagrams for  for $B=0$,   as shown in  FIG.~\ref{pd2}.

\begin{table*}
\begin{tabular}{|l|l|l|l|}
\hline \multicolumn{3}{|c|}{Parameter regimes} & {Ground states $|S^m,S_z^m\rangle$} \\
\hline
&   & $K_e>0$ & $|\frac{N_a-N_b}{2},0\rangle$ \\ \cline{3-4}
 & $C>0$ &$K_e= 0 $ &$|S^m,0\rangle$, $\frac{N_a-N_b}{2} \leq S^m \leq  \frac{N_a+N_b}{2}$ \\ \cline{3-4}
 & & $K_e<0$ &$|\frac{N_a+N_b}{2},0\rangle$\\ \cline{2-4}
 &                           &$Ke>0$ &$|\frac{N_a-N_b}{2},S^m_z\rangle$, $-\frac{N_a-N_b}{2} \leq S^m_z \leq \frac{N_a-N_b}{2}$ \\\cline{3-4}
 $B=0$ & $C = 0$ &$K_e= 0$ & $|S^m,S_z^m\rangle$, $\frac{N_a-N_b}{2} \leq S^m \leq  \frac{N_a+N_b}{2}$, $-S^m \leq S^m_z \leq  S^m $\\\cline{3-4}
 &        & $K_e<0$ &$|\frac{N_a+N_b}{2},S^m_z\rangle$, $-\frac{N_a+N_b}{2} \leq S^m_z \leq \frac{N_a+N_b}{2}$\\ \cline{2-4}
 &       &$K_e > -\frac{N_a C}{N_a+1}$ & $|\frac{N_a-N_b}{2},\pm \frac{N_a-N_b}{2}\rangle$\\ \cline{3-4}
 &$C<0$&$K_e = -\frac{N_a C}{N_a+1}$ &$|\frac{N_a-N_b}{2}, \pm \frac{N_a-N_b}{2}\rangle$ and  $|\frac{N_a+N_b}{2}, \pm \frac{N_a+N_b}{2}\rangle$\\ \cline{3-4}
 & &$K_e < -\frac{N_a C}{N_a+1}$ &$|\frac{N_a+N_b}{2},\pm\frac{N_a+N_b}{2}\rangle$\\ \hline
\end{tabular}
\caption{\label{tableb0} Ground states in various parameter regimes with $B=0$. }
\end{table*}

For $C>0$, $K_e>0$ while $B=0$, the ground state is uniquely $|\frac{N_a-N_b}{2},0\rangle$. In the case of $N_a > N_b$, it is continuously connected with  $|\frac{N_a-N_b}{2}, {\rm Int}(-\frac{B}{2C})\rangle$ in the regime $C > \frac{|B|}{N_a-N_b}$, $K_e>0$ while $B\neq 0$. In the case of $N_a=N_b$, the ground state for $C>0$, $K_e>0$ while $B=0$ becomes $|0,0\rangle$, which is the same one as  for $C>0$ while $K_e>|B|$.

For  $C>0$,  $K_e<0$ while $B=0$, the ground state is uniquely $|\frac{N_a+N_b}{2},0\rangle$. It is continuously connected with $|\frac{N_a+N_b}{2},-{\rm Int}(\frac{B}{2C})\rangle$ in the regime $C> \frac{|B|}{N_a+N_b}$, $K_e < 0$ while $B\neq 0$.

On the boundary $C>0$, $K_e= 0$ while $B=0$, i.e. on the positive $C$ axis, the degenerate ground states are $|S^m,0\rangle$, with $\frac{N_a-N_b}{2} \leq S^m \leq  \frac{N_a+N_b}{2}$, which include the two ground states in the regimes above and below this half axis. In case $N_a > N_b$, these degenerate ground states on positive $C$ axis is continuously connected with $|S^m, {\rm Int}(-\frac{B}{2C})\rangle$, with $\frac{N_a-N_b}{2}  \leq S^m \leq \frac{N_a+N_b}{2}$, on the boundary $K_e=0$, $C \geq \frac{|B|}{N_a-N_b}$ while $B\neq 0$. In case $N_a = N_b$, these ground states are continuously connected with the ground states  $|S^m, {\rm Int}(-\frac{B}{2C})\rangle$, with ${\rm Int}(-\frac{B}{2C})  \leq S^m \leq \frac{N_a+N_b}{2}$, on the boundary $K_e=0$, $C \geq \frac{|B|}{N_a+N_b}$ while $B\neq 0$.

For $C<0$, $K_e > -\frac{N_a C}{N_a+1}$ while $B=0$, the twofold degenerate ground states are  $|\frac{N_a-N_b}{2},\pm \frac{N_a-N_b}{2}\rangle$.  Similarly, for  $C<0$, $K_e < -\frac{N_a C}{N_a+1}$ while $B=0$, the twofold degenerate ground states are $|\frac{N_a+N_b}{2},\pm\frac{N_a+N_b}{2}\rangle$. On the boundary $C<0$, $K_e = -\frac{N_a C}{N_a+1}$ while $B=0$,  the fourfold degenerate ground states  are $|\frac{N_a-N_b}{2}, \pm \frac{N_a-N_b}{2}\rangle$ and $|\frac{N_a+N_b}{2}, \pm \frac{N_a+N_b}{2}\rangle$, just the  union of the ground states above and below this axis.

For $B=0$, on $K_e$ axis, the ground states are always degenerate too. For $C=0$, $K_e>0$ while $B=0$, i.e. on the positive $K_e$ axis, the ground states are $|\frac{N_a-N_b}{2},S^m_z\rangle$, with $-\frac{N_a-N_b}{2} \leq S^m_z \leq \frac{N_a-N_b}{2}$, which include the ground states in the regimes on the right and left of this half axis. This is the ground states discussed previously.  For $B=0$, $C=0$ while $Ke<0$, i.e. on the negative  $K_e$ axis, the ground states are $|\frac{N_a+N_b}{2},S^m_z\rangle$,  with $-\frac{N_a+N_b}{2} \leq S^m_z \leq \frac{N_a+N_b}{2}$, which, again, include the ground states in the regimes on the right and left of this half axis.

At the origin
$B=C= K_e= 0$, the ground state can be any legitimate state, i.e. $|S^m,S_z^m\rangle$, with  $\frac{N_a-N_b}{2} \leq S^m \leq  \frac{N_a+N_b}{2}$ and $-S^m \leq S^m_z \leq  S^m $, or any of their superposition.

Therefore in the case of $B=0$, all the boundaries on $C-K_e$ parameter plane is a phase boundary, on which discontinuities or quantum phase transitions occur. The degenerate ground states on the boundary include the ones in the regimes it divides.

As $B\rightarrow 0$, the boundaries of different $C-K_e$ regime for $B \neq 0$ converge to those of $B=0$. A ground state in each $C-K_e$ regime  on each side of $B=0$ reduces to a ground state in the corresponding $C-K_e$ regime  for $B=0$.  For $C>0$, all the ground states for  $B\neq 0$ are continuously connected with those for $B=0$. For  $B=0$ while $K_e \leq 0$, the ground states  in the regime  $C>0$  can be obtained by substituting $B=0$ in the ground state for  $C>\frac{|B|}{N_a+N_b}$ while $K_e \leq 0$. For $B=0$ while $K_e>0$,  the ground states  in the regime  $C>0$  can be obtained by substituting $B=0$ in the ground state for $C>\frac{|B|}{N_a-N_b}$ while $K_e>0$ in case $N_a >N_b$, and it is the same as the ground state for $C>0$ while $K_e > |B|$ in case $N_a =N_b$.

For $C\leq 0$ while $B=0$, excluding the origin $B=C=K_e=0$, the ground states in each $C-K_e$ regime include ground states on both sides of $B=0$. Hence for $C\leq 0$, there is always a discontinuity or quantum phase transition in the ground state when $B$ switches its sign from positive to negative and vice versa, for any values of $K_e$. When $B$ continuously reduce the magnitude, the ground state remains as the initial one when reaching $B=0$, then the discontinuity in ground state occurs when the sign $B$ becomes opposite to the original.

\section{ground states in terms of boson operators \label{boseexp} }

One can obtain  the expression of any ground state  $|S,S_z\rangle$ in terms of bosonic degrees of freedom, using
\begin{equation}
|S,S_z\rangle =\sum_{S_{bz}=-S_b}^{S_b}  g(S,S_z,S_{bz}) |S_a,S_z-S_{bz}\rangle_a|S_b,S_{bz}\rangle_b, \label{expansion}
\end{equation}
where
$g(S,S_z,S_{bz}) \equiv \langle S_a,S_z-S_{bz},S_b,S_{bz}|S,-S\rangle$ is the Clebsch-Gordan coefficient~\cite{cg}
\begin{widetext}
\begin{equation}
\begin{array}{l}
\displaystyle
g(S,S_z,S_{bz})= \langle S_a,S_z-S_{bz}; S_b,S_{bz}|S_a,S_b,S,S_{z}\rangle
= \displaystyle \left[ \frac{(2S+1)(S_a+S_b-S)!(S_a-S_b+S)!(S_b-S_a+S)!}{(S_a+S_b+S+1)!} \right]^{\frac{1}{2}} \\ \displaystyle
\times [(S_a+S_z-S_{bz})!(S_a-S_{z}+S_{bz})! (S_b+S_{bz})! (S_b-S_{bz})! (S+S_z)! (S-S_z)! ]^{\frac{1}{2}} \\
\times \sum_k (-1)^{k}[k!(S_a+S_b-S-k)!(S_a-S_z+S_{bz}-k)!
(S_b+S_{bz}-k)!(S-S_b+S_z-S_{bz}+k)!(S-S_a-S_{bz}+k)!]^{-1},
\end{array}
\end{equation}
where $k$ is an integer such that the arguments in the factorials are non-negative.  In terms of the particle number in each single particle mode $\alpha =a, b$,
\begin{equation}
|S_{\alpha},S_{\alpha z}\rangle_{\alpha} = |S_{\alpha}+S_{\alpha z}\rangle_{\alpha\uparrow} |S_{\alpha}-S_{\alpha z}\rangle_{\alpha\downarrow}, \label{occunum}
\end{equation}
where
$|n\rangle_{\alpha\sigma}
= \frac{1}{\sqrt{n!}}(\alpha^{\dagger}_{\sigma})^{n}|0\rangle,$
($\sigma = \uparrow, \downarrow$).

Therefore,
\begin{equation}
|S,S_z\rangle =\sum_{S_{bz}=-S_b}^{S_b}  f(S,S_z,S_{bz}) (a^{\dagger}_{\uparrow})^{S_a+S_z-S_{bz}} (a^{\dagger}_{\downarrow})^{S_a-S_z+S_{bz}}
(b^{\dagger}_{\uparrow})^{S_b+S_{bz}} (b^{\dagger}_{\downarrow})^{S_b-S_{bz}}|0\rangle, \label{expansion2}
\end{equation}
with $$f(S,S_z,S_{bz})=g(S,S_z,S_{bz})[(S_a+S_z-S_{bz})!(S_a-S_z+S_{bz})!
(S_b+S_{bz})!(S_b-S_{bz})!]^{-1/2}.$$

In some special cases, to which most of our ground states belong, this expression can be simplified as the following.

\subsection{$|S, \pm S\rangle$}

For $S_z=-S$, we have
\begin{equation}
g(S,-S,S_{bz})  = (-1)^{S_{b}+S_{bz}} \left[ \frac{(2S+1)!(S_a+S_b-S)!(S_a+S_{bz}+S)!(S_b-S_{bz})!}{(S_a+S_b+S+1)!(S_a-S_b+S)!
(S_a-S-S_{bz})!(S_b+S_{bz})!} \right]^{\frac{1}{2}}.
\end{equation}
Therefore
\begin{equation}
f(S,-S,S_{bz})=(-1)^{S_b+S_{bz}} \Gamma_1 \frac{(S_a+S_b-S)!}{(S_a-S-S_{bz})!(S_b+S_{bz})!},
\label{ga}
\end{equation}
where
$\Gamma_1\equiv\left\{\frac{(2S+1)!}
{(S_{max}+S+1)!(S_{max}-S)!(S_{min}+S)!}\right\}^{1/2}$. Thus
\begin{equation}
|S,-S\rangle= \Gamma_1 (a^\dagger_\downarrow)^{S+\frac{N_a-N_b}{2}}(b^\dagger_\downarrow)^{S-
\frac{N_a-N_b}{2} }(a^\dagger_\uparrow b^\dagger_\downarrow-a^\dagger_\downarrow b^\dagger_\uparrow)^{\frac{N_a+N_b}{2}-S}|0\rangle.
\end{equation}

Similarly, for $S_z=S$, we have
\begin{equation}
\displaystyle
g(S,S,S_{bz})  = (-1)^{S_a-S+S_{bz}} \left[ \frac{(2S+1)!(S_a+S_b-S)!(S_a-S_{bz}+S)!(S_b+S_{bz})!}{(S_a+S_b+S+1)!(S_a-S_b+S)!
(S_a-S+S_{bz})!(S_b-S_{bz})!} \right]^{\frac{1}{2}}.
\end{equation}
Therefore
\begin{equation}
f(S,S,S_{bz})=(-1)^{S_a-S+S_{bz}}\Gamma_1 \frac{(S_a+S_b-S)!}{(S_a-S-S_{bz})!(S_b+S_{bz})!}.
\label{ga1}
\end{equation}
Thus
\begin{equation}
|S,S\rangle= \Gamma_1 (a^\dagger_\uparrow)^{S+\frac{N_a-N_b}{2}}(b^\dagger_\uparrow)^{S-
\frac{N_a-N_b}{2}}(a^\dagger_\uparrow b^\dagger_\downarrow-a^\dagger_\downarrow b^\dagger_\uparrow)^{\frac{N_a+N_b}{2}-S}|0\rangle.
\end{equation}

It is clear that in $|S,\pm S\rangle$, there are $\frac{N_a+N_b}{2}-S$ interspecies singlet pairs, plus $S+\frac{N_a-N_b}{2}$ $a$-atoms and $S-\frac{N_a-N_b}{2}$ $b$-atoms, each of which has $z$-component spin polarized as $\pm 1$. This is fully consistent with discussions in terms of the generating function method~\cite{shi4}.

\subsection{$|\frac{N_a-N_b}{2},S_z\rangle$}

For $S=S_{min}$, we have
\begin{equation}
g(S_{min},S_z,S_{bz})=(-1)^{S_{bz}} \left[ \frac{(2S_a-2S_b+1)!(2S_b)!(S_a+S_z-S_{bz})!(S_a-S_{z}+S_{bz})!}
{(2S_a+1)!(S_b+S_{bz})!(S_b-S_{bz})!
(S_a-S_b+S_{z})!(S_a-S_b-S_{z})!} \right]^{\frac{1}{2}}.
\end{equation}
Therefore
\begin{equation}
f(S_{min},S_z,S_{bz})=\Gamma_2 (-1)^{S_{bz}}\frac{(2S_b)!}{(S_b+S_{bz})!(S_b-S_{bz})!},
\end{equation}
where $\Gamma_2 \equiv \left\{\frac{(2S_a-2S_b+1)!}{(2S_a+1)!(S_a-S_b+S_z)!(S_a-S_b-S_z)!(2S_b)!}\right\}^{1/2}$.
Thus
\begin{equation}
|\frac{N_a-N_b}{2},S_z\rangle=\Gamma_2(a^\dagger_\uparrow)^{\frac{N_a-N_b}{2}
+S_z}(a^\dagger_\downarrow)^{\frac{N_a-N_b}{2}-S_z}(a^\dagger_\uparrow b^\dagger_\downarrow- a^\dagger_\downarrow b^\dagger_\uparrow)^{N_b}|0\rangle,
\end{equation}
as obtained in \cite{shi1}.

\subsection{$|\frac{N_a+N_b}{2},S_z\rangle$}

For $S=S_{max}$, we have
\begin{equation}
g(S_{max},S_z,S_{bz})=
\left\{\frac{(2S_b)!(2S_a)!(S_a+S_b+S_z)!(S_a+S_b-S_z)!}
{(2S_a+2S_b)!(S_a+S_z-S_{bz})!(S_a-S_z+S_{bz})!(S_b+S_{bz})!(S_b-S_{bz})!}
\right\}^{1/2}.
\end{equation}
Therefore
\begin{equation}
|\frac{N_a+N_b}{2},S_z\rangle=\Gamma_3
\sum_{S_{bz}=-S_b}^{S_b} c(S_{bz}) (a^{\dagger}_{\uparrow})^{S_a+S_z-S_{bz}} (a^{\dagger}_{\downarrow})^{S_a-S_z+S_{bz}}
(b^{\dagger}_{\uparrow})^{S_b+S_{bz}} (b^{\dagger}_{\downarrow})^{S_b-S_{bz}}|0\rangle,
\end{equation}
where $\Gamma_3=
\left\{\frac{(2S_b)!(2S_a)!(S_a+S_b+S_z)!(S_a+S_b-S_z)!}{(2S_a+2S_b)!}
\right\}^{1/2},$ and $c(S_{bz})=[(S_a+S_z-S_{bz})!(S_a-S_z+S_{bz})!(S_b+S_{bz})!
(S_b-S_{bz})!]^{-1},$ which is nonzero only if the all the arguments of factorials are non-negative.

Especially, when $S_z=\pm(S_a+S_b)$, the state is a direct product of states of individual species.
\begin{equation}
|\frac{N_a+N_b}{2},\pm\frac{N_a+N_b}{2}\rangle = |\frac{N_a}{2}, \pm \frac{N_a}{2}\rangle_a|\frac{N_b}{2},\pm \frac{N_b}{2}\rangle_b,
\end{equation}
where
\begin{equation}
|\frac{N_{\alpha}}{2}, \pm \frac{N_{\alpha}}{2}\rangle_{\alpha} = \frac{1}{\sqrt{N_{\alpha}!}}(\alpha^{\dagger}_{\pm})^{N_{\alpha}}|0\rangle,
\end{equation}
with $\alpha^{\dagger}_{\pm}$ representing   $\alpha^{\dagger}_{\uparrow}$ and   $\alpha^{\dagger}_{\downarrow}$,    is a ferromagnetic state of species $\alpha$.

\end{widetext}

\section{Entanglement  \label{prop}}

In any state $|S,S_z\rangle$, the two species are entangled unless $S=S_a+S_b$ and $S_z=\pm S$. In general, we can call a state polarized entangled BEC when $S_z \neq 0$ and there is interspecies entanglement.
For any state $|S,S_z\rangle$, we calculate the entanglement between the two species,  which is quantified as the von Neumann entropy of the reduced density matrix of either species.   One may consider the entanglement between the occupation number of each of the four single-particle modes and the rest of the
system~\cite{shientropy}. Alternatively one simply considers the total the spins of the two species.  Both approaches yield the same entanglement entropy ${\cal E}$,
\begin{equation}
{\cal E} =
-\sum_{m=-S_b}^{S_b} g^2(S,S_z,m)\log_{N_b+1}g^2(S_z,m).
\end{equation}
The reason why both approaches lead to the same result,  for any $|S,S_z\rangle$,  is that the expansion is always automatically Schmidt decomposition with coefficients $g(S,S_z,m)$, no matter it is in terms of the spin basis states of the two species, as given in (\ref{expansion}), or in terms of the occupation numbers of each species and each pseudospin state, as obtained by substituting (\ref{occunum}) to (\ref{expansion}).

In the following, we focus on the maximally polarized state
$|S,\pm S\rangle$, because of the following reasons. First, most of our ground states are in this form. Second, in case the ground states are degenerate with respect to $S_z$, an infinitesimal symmetry breaking perturbation proportional to $S_z$ picks out a maximally polarized state as the new ground state. Clearly, the entanglement entropies of $|S,S\rangle$ and of $|S,-S\rangle$ are equal, because $g(S,m)=(-1)^{S_{min}-S} g(-S,-m)$.

For simplicity, we reduce our focus to the case  of $N_a=N_b=N$, in which  $|S,-S\rangle=\Gamma_1 (a^\dagger_\downarrow b^\dagger_\downarrow)^{S}(a^\dagger_\uparrow b^\dagger_\downarrow-a^\dagger_\downarrow b^\dagger_\uparrow)^{N-S}|0\rangle$ and
$|S,S\rangle=\Gamma_1 (a^\dagger_\uparrow b^\dagger_\uparrow)^{S}(a^\dagger_\uparrow b^\dagger_\downarrow-a^\dagger_\downarrow b^\dagger_\uparrow)^{N-S}|0\rangle$.

\begin{figure*}
\scalebox{0.9}[0.9]{\includegraphics{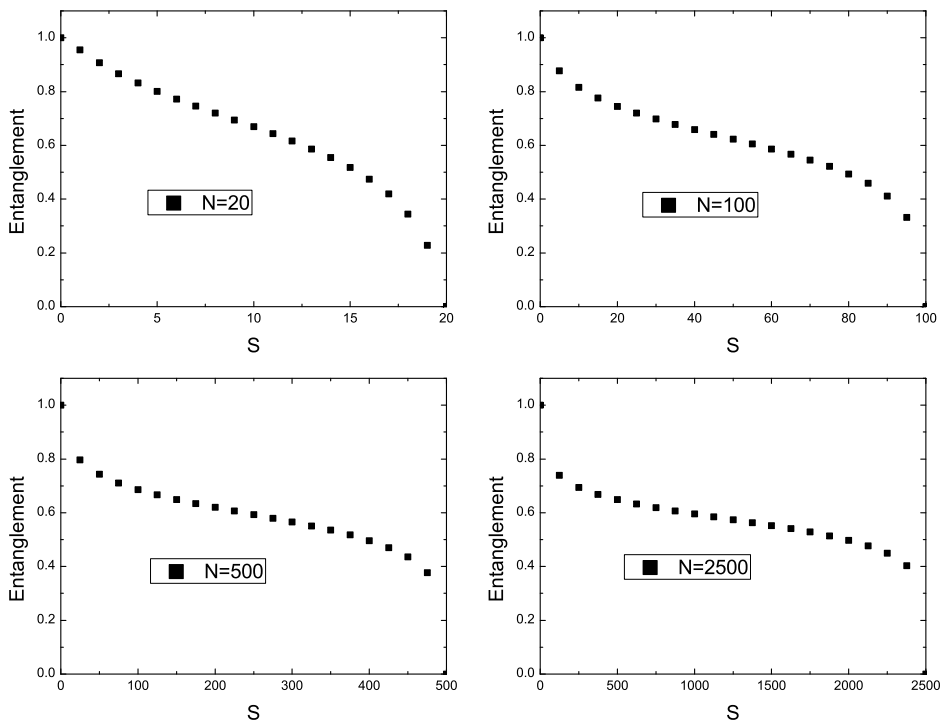}}
\caption{Entanglement entropy of the interspecies entanglement in state $|S,\pm S\rangle$, with $N_a=N_b=N$, as a function of $S$. \label{ENT}}
\end{figure*}

The calculation results are
shown in FIG.~\ref{ENT} for four values of $N$. It can be seen that for a given $N$,  entanglement always decreases when $S$ increases.

We know
\begin{equation}
N_{\alpha\sigma} = N_{\alpha}/2 + \eta_{\sigma} S_{\alpha z},
\end{equation}
with  $\eta_{\uparrow} =1 $ and  $\eta_{\downarrow} =-1$.  Therefore, for any state, the average number of $\alpha$-particles with spin $\sigma$ is \begin{equation}
\langle N_{\alpha \sigma} \rangle= \frac{N}{2}+\eta_{\sigma}
M_{\alpha 1},
\end{equation}
where
\begin{equation}
M_{\alpha 1}\equiv \langle S_{az}\rangle,
\end{equation}
Its fluctuation is
\begin{equation}
\Delta N_{\alpha\sigma} \equiv \sqrt{\langle N_{\alpha\sigma}^2\rangle  -( \langle N_{\alpha\sigma}\rangle)^2 } = (M_{\alpha 2}-M_{\alpha 1}^2)^{1/2},
\end{equation}
where
\begin{equation}
M_{\alpha 2}\equiv \langle S^2_{az}\rangle.
\end{equation}

\begin{figure*}
\scalebox{0.9}[0.9]{\includegraphics{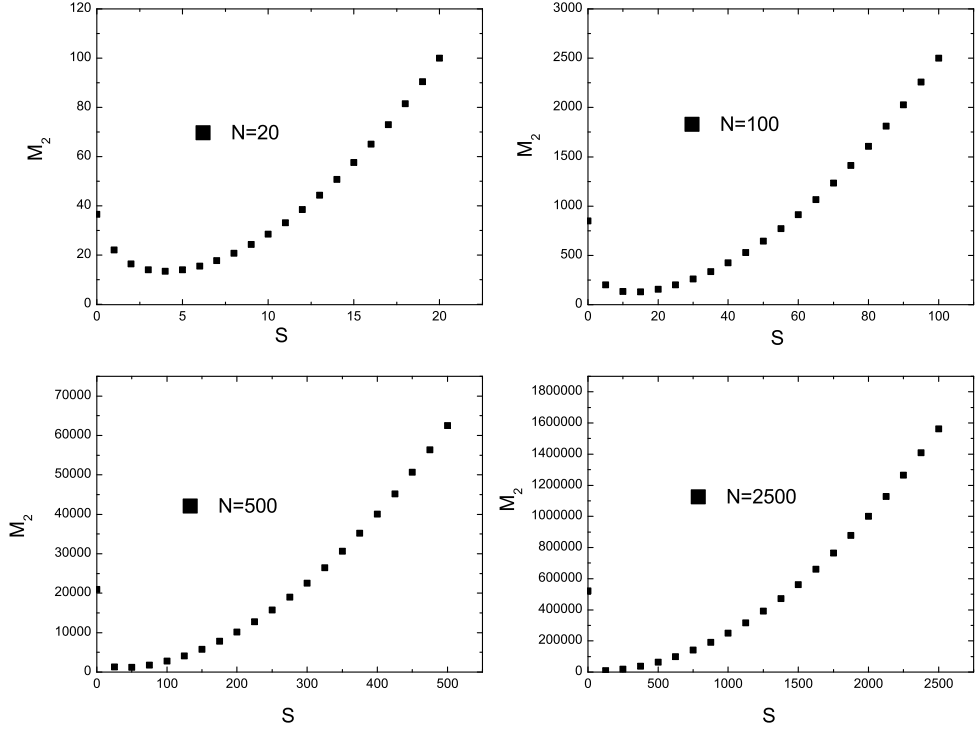}}
\caption{ $M_2 \equiv \langle S_{\alpha z}^2 \rangle$, for state $|S,\pm S\rangle$,  as a function of $S$. Each plot is for a given $N$ as indicated.  \label{SAZ2}}
\end{figure*}

\begin{figure*}
\scalebox{0.9}[0.9]{\includegraphics{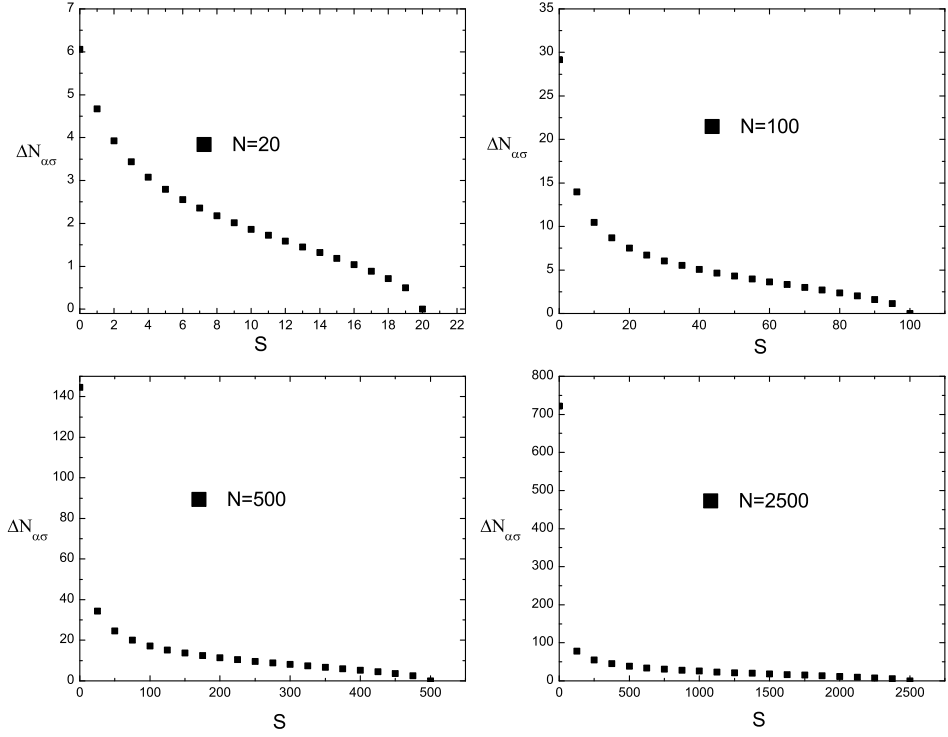}}
\caption{The fluctuation $\Delta N_{\alpha\sigma}$ , for state $|S,\pm S\rangle$,  as a function of $S$. Each plot is for a given $N$ as indicated.  \label{SAZ2r}}
\end{figure*}

For $|S,\eta_S S\rangle$, with $\eta_S =\pm 1$, it is obtained that
$M_{a1}= M_{b1} =\eta_S \frac{S}{2}$. Hence the total polarization is $\eta_SS$. It is also obtained that for this state, $M_{a 2}= M_{b2}$, denoted as
$M_2$, which depends on $S$ and $N$ in a more complicated manner,  as shown in FIG.~\ref{SAZ2}. For this state, one also has $\Delta N_{a\uparrow}=\Delta N_{a\downarrow}=\Delta N_{b\uparrow}=\Delta N_{b\downarrow}$, which is shown in FIG~\ref{SAZ2r}. One can see that the dependence of the fluctuation $\Delta N_{\alpha\sigma}$ on  $S$ has a  trend  similar to the entanglement entropy. Indeed, the fluctuation  $\Delta N_{\alpha\sigma}$ is also a characterization of the interspecies entanglement~\cite{shi1}.

Both $M_{\alpha 1}$ and $M_2$ enter the Gross-Pitaevskii-like equation governing the single-particle orbital wave function $\phi_{\alpha\sigma}(\vec{r})$ under the ground state $|S,\eta_S S\rangle$, with $S <N$,
\begin{widetext}
\begin{equation}
\begin{array}{c}
\displaystyle
\left(\frac{N}{2}+\eta_{\sigma}\eta_S\frac{S}{2}\right)
\left[-\frac{\hbar^2}{2m_\alpha}\nabla^2
+U_{\alpha\sigma}(\mathbf{r}_\alpha)\right]\phi_{\alpha\sigma}+
\Big[\frac{N^2}{4}+\eta_{\sigma}\eta_S\frac{(N-1)S}{2}+M_2-\frac{N}{2}\Big]
g^{(\alpha\alpha)}_{\sigma\sigma}|\phi_{\alpha\sigma}|^2
\phi_{\alpha\sigma} \nonumber \\ \displaystyle +\Big(\frac{N^2}{4}-M_2\Big)
g^{(\alpha\alpha)}_{\sigma\bar{\sigma}}|\phi_{\alpha\bar{\sigma}}|^2
\phi_{\alpha\sigma}+\Big(\frac{N^2}{4}+
\eta_\sigma\eta_S\frac{NS}{2}+\frac{S^2}{2}-M_2\Big)
g^{(\alpha\beta)}_{\sigma\sigma}|\phi_{\beta\sigma}|^2\phi_{\alpha\sigma}+
\Big(\frac{N^2}{4}-\frac{S^2}{2}+M_2\Big)
g^{(\alpha\beta)}_{\sigma\bar{\sigma}}|\phi_{\beta\bar{\sigma}}|^2
\phi_{\alpha\sigma}\nonumber\\ \displaystyle  + g_e\Big(M_2-\frac{N^2}{4}-
\frac{N}{2}+\frac{S}{2}\Big)
\phi^*_{\beta\bar{\sigma}}\phi_{\beta\sigma}\phi_{\alpha\bar{\sigma}}=
\mu_{\alpha\sigma}\Big(\frac{N}{2}+\eta_\sigma\eta_S\frac{S}{2}\Big)
\phi_{\alpha\sigma}. \label{gp1}
\end{array}
\end{equation}
\end{widetext}
It is obtained by requiring the variation of the energy with respect to $\phi_{\alpha\sigma}$ to be minimum, under the constraint $\int  |\phi_{\alpha\sigma}|^2  d\mathbf{r} =1 $.  Note that now the energy $\langle {\cal H} \rangle$, where ${\cal H}$ is as given in (\ref{h}), is treated as a functional of $\phi_{\alpha\sigma}$. The effective coefficients $\epsilon_{\alpha\sigma}$ and $K$'s are now functionals of $\phi_{\alpha\sigma}$. In the limit $S=0$, the is equation reduces to that under the singlet ground state  previously studied.

It can be seen that interspecies entanglement significantly affects the Gross-Pitaevskii equations, noting $M_2=(\Delta N_{\alpha\sigma})^2+S^2/4$.

When $S=N$, the ground states $|N, \eta_N N\rangle$  is disentangled between the two species, consequently there are only $\phi_{\alpha\sigma}$ ($\alpha =a,b$) with $\sigma$ specified by  $\eta_{\sigma}\eta_N =1$. For $|N,N\rangle = (1/N!) {a_{\uparrow}^{\dagger}}^N  {b_{\uparrow}^{\dagger}}^N|0\rangle $, there are only  $\phi_{a\uparrow}$ and $\phi_{b\uparrow}$. For $|N,-N\rangle = (1/N!) {a_{\downarrow}^{\dagger}}^N  {b_{\downarrow}^{\dagger}}^N|0\rangle$,  there are only  $\phi_{a\downarrow}$ and $\phi_{b\downarrow}$. For these two states, $M_2=N^2/4$, and thus (\ref{gp1}) reduces to
\begin{widetext}
\begin{equation}
\begin{array}{c}
\displaystyle
\left[-\frac{\hbar^2}{2m_\alpha}\nabla^2
+U_{\alpha\sigma}(\vec{r}_\alpha)\right]\phi_{\alpha\sigma}+(N-1)
g^{(\alpha\alpha)}_{\sigma\sigma}|\phi_{\alpha\sigma}|^2
\phi_{\alpha\sigma}+N
g^{(\alpha\beta)}_{\sigma\sigma}|\phi_{\beta\sigma}|^2\phi_{\alpha\sigma}=
\mu_{\alpha\sigma}
\phi_{\alpha\sigma}. \label{gp2}
\end{array}
\end{equation}
\end{widetext}
which is just Gross-Pitaevskii equations for the well studied two-component BEC.

\section{Summary \label{summary} }

To summarize, we have considered a more general case of the many-body Hamiltonian of a mixture of two species of pseudospin-$\frac{1}{2}$ bosons with interspecies spin-exchange interaction, with three effective parameters as given in (\ref{hamiltonianall}).  We have determined the ground states in all the regimes of these three parameters. Discontinuities of the ground states  occur in some parameter boundaries.  These  are first order quantum phase transitions, as $S^2$ or $S_z$ have discontinuities in crossing the boundaries.

The two species are entangled in all the ground states except $|\frac{N_a+N_b}{2},\pm \frac{N_a+N_b}{2}\rangle$, which is the ground states when $K_e$ and $C$ are both negative or  sufficiently small positive, as shown in the phase diagrams.
Very interestingly, when $N_a=N_b$, the ground state is the global singlet state $|0,0\rangle$ in the wide regime in which $K_e $ is greater than the larger one between $\frac{|B|-NC}{N+1}$ and $|B|$.  This result confirms the previous arguments based on numerical estimation  and perturbative analysis that the singlet ground state and the  interspecies entanglement persist in a wide parameter regime~\cite{shi1,shi2}.

We have calculated the properties of $|S,\pm S\rangle$ in details. For a given $N$,  the polarization  is equal to $\pm S$. The entanglement decreases with as $S$ increases. It reaches the maximal value $1$ as $S=0$, and reaches the minimal value $0$ as $S=N$. The fluctuation of the particle number in either pseudospin state of each species has a similar behavior as the entanglement, reaching the maximal $\sqrt{N(N+2)/12}$ as $S=0$~\cite{shi1}, and reaching the minimum $0$ as $S=N$.
The interspecies entanglement and polarization significantly affect the Gross-Pitaevskii equations governing the orbital wave functions associated with each pseudospin state of each species. This is a very interesting interplay between spin and orbital degrees of freedom. More phenomenology beyond that of the two-component BEC and can be experimentally observed is under investigation.

\appendix

\section{$S^m$ and $S^m_{z}$
in all parameters regimes with $B\neq 0$ \label{bneq0} }

In this and next  appendices, we find $S^m$ in all regimes of the parameters. Neighboring regimes with a same ground state  can be combined, as described in the main text. We consider $B \neq 0$  in this Appendix.

\subsection{$B \neq 0$, $C>0$ \label{bb}}

Were there not be the constraint (\ref{range2}) on $S_z$, the minimum of the second term of (\ref{e2}) would be then at $S_z= S_{z0}$, where
$$S_{z0} \equiv {\rm Int}(- B/2C)$$ is the integer closest to $- B/2C$.   But one should, of course,  consider the constraint (\ref{range2}), with the bounds $\pm S$ determined by the sign of $K_e$.

\subsubsection{$B \neq 0$, $C>0$, $K_e<0$ \label{bb1}}

Then $S^m=S_{max}$, under which $S_z$ can vary in the largest possible range. $S_{z}^{m}$ is determined by comparing $S_{z0}$ with the bounds $\pm S_{max}$.  If $-S_{max} \leq S_{z0} \leq S_{max}$, that is, $C \geq \frac{|B|}{N_a+N_b}$, then we have  $S^m_z={\rm Int}(-\frac{B}{2C})$. If $S_{z0} \leq -S_{max}$, that is,  $C \leq \frac{B}{N_a+N_b}$, then $S^m_z=-S_{max}$.  If $S_{z0} \geq S_{max}$, that is, $C \leq -\frac{B}{N_a+N_b}$, then $S^m_z=S_{max}$.

\subsubsection{$B \neq 0$, $C>0$, $K_e=0$  \label{bb2}}

Now there is no $K_e$ term. However, $-S_{max} \leq S \leq S_{max}$ is the largest possible range of $S$. If $C \leq \frac{|B|}{N_a+N_b}$ , we have $S^m_z=-{\rm sgn}(B)S_{max}$, which forces $S^m=S_{max}$. If $\frac{|B|}{N_a+N_b} \leq C \leq \frac{|B|}{N_a-N_b}$, then  $S^m_z={\rm Int}(-\frac{B}{2C})$ and thus $|{\rm Int}(-\frac{B}{2C})| \leq S^m \leq \frac{N_a+N_b}{2}$. If $C \geq \frac{|B|}{N_a-N_b}$, then  $S^m_z={\rm Int}(-\frac{B}{2C})$ and $\frac{N_a-N_b}{2} \leq S^m \leq \frac{N_a+N_b}{2}$.

\subsubsection{$B \neq 0$, $C>0$, $K_e>0$  \label{bb3}}

The minimum of the  first term of (\ref{e2}) is at $S=S_{min}$, which constrains the range of $S_z$. As $S_z$ term is positive, $S^m$ may not be $S_{min}$.  Nevertheless, if  $-S_{min} \leq S_{z0} \leq S_{min}$, that is,  $C \geq \frac{|B|}{N_a-N_b}$, then $S^m=S_{min}$ and $S^m_z=S_{z0}$.  If $ S_{z0} \leq -S_{min}$ or $ S_{z0} \geq S_{min}$, that is,  if $C\leq \frac{|B|}{N_a-N_b}$, then we have $S^m_z=-{\rm sgn}(B)S^m$.

$S_m$ will be determined below altogether for various parameter regimes with $S_z^m = -{\rm sgn}(B) S^m$. The present condition $C>0$ and $K_e>0$  overlaps only with the subcase $K_e > -C$ in Appendix~\ref{szs}. Hence there are three possibilities under the condition $C\leq \frac{|B|}{N_a-N_b}$.

\subsection{$B \neq 0$, $C=0$   \label{c}}

In this case, in order that $B S_z$ is smallest, we have $S_z^m = -{\rm sgn}(B) S^m$, where ${\rm sgn}(B)$ represents the sign of $B$, $S^m$ is the value of $S$ which minimizes $E$. Using the result of  Appendix~\ref{szs}, we know that there are three subcases. For $K_e=0$, $S^m=S_{max}$. For $K_e >0$, the result of  applies. For $K_e < 0$, the result of  applies.

\subsection{$B \neq 0$, $C<0$  \label{d} }

With $C<0$, the $S_z$-dependent term in the energy (\ref{e2}) is minimized always at $S_{z}^m=-{\rm sgn}(B)S_{m}$. This can be seen by representing this term as a parabola opening downward, with  $S_{z0}$ the maximal point. All three subcases of Appendix apply under the condition $C<0$.

\subsection{The cases with $B\neq 0$ and $S_z=-{\rm sgn}(B)S$ \label{szs}}

As discussed above, for the case of $B\neq 0$ and $C \leq 0$, as well as the case of $B\neq 0$, $C>0$ while $K_e >0$, we have $S_z=-{\rm sgn}(B)S$, therefore $E$ can  be expressed to

\begin{eqnarray}
E & = & \left(K_e+C \right)S^2+\left(K_e-|B|\right)S \label{mini1}  \\
& = & (K_e+C)(S-S_0)^2+ E_0'' \label{mini},
\end{eqnarray}
where
\begin{eqnarray}
S_0 \equiv \frac{|B|-K_e}{2(K_e+C)},
\end{eqnarray}
is the the extreme point of the parabola $(K_e+C)(S-S_0)^2$, $E_0''$ is independent of  $S$ and $S_z$.

\subsubsection{$K_e=-C$}

This condition have overlap with $C \leq 0$, while do not overlaps with the condition $C>0$ and $K_e >0$. Then  $E=(K_e-|B|)S+const$. Therefore we know the following subcases, which .

If $K_e =-C < |B|$,  $S^m=S_{max}$.

If $K_e = -C > |B|$, $S^m=S_{min}$.

If $K_e = -C = |B|$, $S^m$ is any any integer in the range $S_{min}\leq S^m \leq S_{max}$.

\subsubsection{$K_e > -C$}

This condition overlaps with both the condition $C\leq 0$ and the condition  $C>0$ and $K_e >0$.

Now that $K_e+C >0$, the parabola $(K_e+C)(S-S_0)^2$ open upwards. There are several subcases depending on the range of $S_0$, as one can conceive by considering the position of $S_0$ related to the range (\ref{range1}).  (i) If $S_0 \leq S_{min}$, which requires $ K_e \geq  \frac{|B|-(N_a-N_b)C}{N_a-N_b+1}$, then $S^m=S_{min}$. (ii) If  $S_{min} \leq S_0 \leq S_{max}$, which requires $\frac{|B|-(N_a+N_b)C}{N_a+N_b+1} \leq  K_e \leq \frac{|B|-(N_a-N_b)C}{N_a-N_b+1}$, which is a subset of $-C < K_e < |B|$, then $S^m=S_0$. (iii) If $S_0\geq S_{max}$, which requires  $-C < K_e \leq  \frac{|B|-(N_a+N_b)C}{N_a+N_b+1}$, then $S^m=S_{max}$.

\subsubsection{$K_e < -C$ \label{szs3}}

This condition overlaps with the condition $C\leq 0$, while do not overlap with the condition  $C>0$ and $K_e >0$.

With $K_e+C <0$, the coefficient in (\ref{mini}) is negative. The parabola $(K_e+C)(S-S_0)^2$ open downwards.

We find that $S^m =S_{max}$ for $ K_e < \frac{|B|-N_aC}{N_a+1}$, while $S^m=S_{min}$ for $\frac{|B|-N_aC}{N_a+1} < K_e < -C $, and $S^m=S_{max}$ and $S^m=S_{min}$ are two degenerate solutions for $K_e < \frac{|B|-N_aC}{N_a+1}$. This result is obtained by considering
the position of $S_0$ relative to the range (\ref{range1}), as the following.  (i) $ S_0 < S_a$, which requires
$K_e < \frac{|B|-N_aC}{N_a+1}$,
then $S^m=S_{max}$. (ii) $S_0= S_a$, which requires $K_e = \frac{|B|-N_aC}{N_a+1}$, there are two degenerate solutions $S^m =S_{max}$ and $S^m=S_{min}$.
(iii) When $S_0 > S_a $, which requires
$K_e > \frac{|B|-N_aC}{N_a+1}$,
then $S^m=S_{min}$.

\section{$S^m$ and $S^m_{z}$ in all parameters regimes with $B=0$ \label{beq0} }

Now we look at the special situation of $B=0$.

\subsection{$B=0$, $C>0$}

Then $S_z^m =0$ in order to minimize $CS_z^2$. Subsequently, it is easy to see that  $S^m =S_{min}$ if
$K_e > 0$,  $S^m =S_{max}$ if  $K_e < 0$, and $S^m$ can be any legitimate $S$, i.e. $S_{min} \leq S^m \leq S_{max}$, if $K_e=0$.

\subsection{$B=0$, $C=0$}

There is no $S_z$ term in $E$. Hence the values of $S^m$ are the same for $C>0$, while $S^m_z$ can be any value in the legitimate ranges. If  $K_e > 0$,  $S^m =S_{min}$ while  $-S_{min} \leq S_z^m \leq S_{min}$. If $K_e < 0$,  $S^m =S_{max}$  while   $-S_{max} \leq S_z^m \leq S_{max}$. If $B=C=K_e=0$, then  $S_m$ and $S_z^m$ are arbitrary values within the ranges (\ref{range1}) and (\ref{range2}).

\subsection{$B=0$, $C<0$}

Then we have  $S_z= \pm S$ in order to minimize $E$, consequently $E$ can be written as Equations (\ref{mini1}) and (\ref{mini}), with $B=0$ and thus $S_0=-\frac{K_e}{2(K_e+C)}$. The discussions in App.~\ref{szs} apply with $B$ set to $0$. But as $B=0$ causes both simplification and constraint. So we give details in the following.

\subsubsection{$B=0$, $C<0$, $K_e=-C$}

Now  $E=K_eS+const$. If $K_e =-C < 0$,  $S^m=S_{max}$,  $S_z^m=\pm S_{max}$.
If $K_e = -C > 0$, $S^m=S_{min}$,  $S_z^m=\pm S_{min}$.

\subsubsection{$B=0$, $C<0$, $K_e > -C$}

Now that $K_e+C >0$, the parabola $(K_e+C)(S-S_0)^2$ open upwards. On the other hand now $K_e>-C>0$, thus  $S_0=-\frac{K_e}{K_e+C} <0 $. Therefore
$S^m=S_{min}$, $S_z^m=\pm S_{min}$.

\subsubsection{$B=0$, $C<0$, $K_e < -C$}

With $K_e+C <0$, the parabola $(K_e+C)(S-S_0)^2$ open downwards. We can directly use the result in App.~\ref{szs}, setting $B=0$, to know that
$S^m =S_{max}$ for $ K_e < -\frac{N_aC}{N_a+1}$, while $S^m=S_{min}$ for $-\frac{N_aC}{N_a+1} < K_e < -C $, and $S^m=S_{max}$ and $S^m=S_{min}$ are two degenerate solutions for $K_e = -\frac{N_aC}{N_a+1}$.

\acknowledgments

This work was supported by the National Science Foundation of China (Grant No. 11074048), the Shuguang Project (Grant No. 07S402) and the Ministry of Science and Technology of China (Grant No. 2009CB929204).

\end{document}